\documentclass[pra,preprint,showpacs]{revtex4} 

\usepackage{amsmath}  
\usepackage{amsfonts} 
\usepackage{graphicx} 

\usepackage{dsfont} %For the identity operator

\newcommand{\be}{\begin{equation}}
\newcommand{\ee}{\end{equation}}

\newcommand{\sdual}{\check s}

\newcommand{\lexpval}{{\mathbb E}_s^\lambda [A]}
\newcommand{\lexpvalz}{{\mathbb E}^{\lambda=0} [A]}
\newcommand{\wexpval}{{\mathbb E}_s^w [A]}
\newcommand{\wexpvalc}{{\mathbb E}_{\scombined}^w [A]}
\newcommand{\wexpvalj}{{\mathbb E}_{s_j}^w [A]}
\newcommand{\lexpvalp}{{}_{\sdual}{\mathbb E}_s^\lambda [A]}
\newcommand{\wexpvalp}{{}_{\sdual}{\mathbb E}_s^w [A]}
\newcommand{\wexpvalpp}{{}_{\sdual}{\mathbb E}_s^{w'} [A]}
\newcommand{\qxpval}[1]{\langle #1 \rangle}
\newcommand{\states}{\mathcal S}
\newcommand{\bstates}{\mathcal S'}
\newcommand{\dstates}{\check {\mathcal S}}
\newcommand{\idop}{\mathds 1} 
\newcommand{\hilb}{\mathcal H}
\newcommand{\hilba}{{\mathcal H}_{\text{aux}}}

\newcommand{\saux}{s_{\text{aux}}}

\newcommand{\khat}{\hat K}

\newcommand{\ud}{\text d}
\newcommand{\scombined}{s\cdot\sdual} %Something better???

\newcommand{\distmeas}{\mathcal F}
\newcommand{\distmeasf}{f}
\newcommand{\measmnts}{\mathcal M}

\DeclareMathOperator{\tr}{tr}
\DeclareMathOperator{\re}{Re}
\DeclareMathOperator{\im}{Im}
\DeclareMathOperator{\ndo}{End}
\DeclareMathOperator{\sgn}{sgn}

\begin{document}

\title{Disturbance in weak measurements and the difference between quantum
and classical weak values}

\author{Asger C.~Ipsen}
\email{asgercro@nbi.dk} 
\affiliation{Niels Bohr Institute, University of Copenhagen, Blegdamsvej 17, 2100 Copenhagen {\O}, Denmark}

\date{\today}

\begin{abstract}
The role of measurement induced disturbance in weak measurements is of central
 importance for the interpretation of the weak value. 
Uncontrolled disturbance can interfere with the postselection process and make
the weak value dependent on the details of the measurement process.
Here we develop the concept of a generalized weak measurement  for 
classical and quantum mechanics. The two cases appear remarkably similar, but we
point out some important differences. 
A priori it is not clear what the correct notion of disturbance should be in
the context of weak measurements.
We consider three different notions and get three different results:
(1) For a `strong' definition of disturbance, we find that weak measurements are
 disturbing. (2) For a weaker definition we find that a  general class of weak
measurements are non-disturbing, but that one gets weak values which depend on the 
measurement process. (3) Finally, with respect to an operational definition of 
the `degree of disturbance', we find
that the AAV weak measurements are the least disturbing,
but that the disturbance is always non-zero.
\end{abstract}


\maketitle 
\section{Introduction}
It has been proposed that weak values\cite{AAV88,AV90,APRV93} could serve  
as an operational definition of the expectation values
of observables in the intermediate time between the preparation of a system
and a postselection on the final state of the system. If one were to employ
standard (strong) measurements between the preparation and postselection the
inevitable disturbance caused by the measurement (see e.g.~the ``No Information
 Gain Without Disturbance'' theorem of Ref.~\cite{Bus09}) would interfere with
the postselection. The central idea of weak measurements is to avoid the issue
of disturbance by making the interaction between the measurement apparatus and the
system during the intermediate measurement arbitrarily small. 
This was expressed explicitly in Ref.~\cite{APRV93} as:
\begin{quote}
   ...the [weak] measurements hardly disturb the ensemble, and therefore
   they characterize the ensemble during the whole intermediate time [between preparation
   and postselection].
\end{quote}
Motivated by recent work\cite{DJ12B,FC14} we explore role of disturbance and
the relation between classical and quantum mechanical weak measurements.

For a classical systems the expectation value of some observable between a 
preparation and a postselection has a perfectly unambiguous meaning, and if one 
applies the weak measurement procedure with a non-disturbing measurement one recovers
the `correct' expectation value. However, if one allows the intermediate measurement
to disturb the system, \emph{even if this disturbance goes to zero along with the interaction
strength}, one can get results that deviates from this value (see Section \ref{sec:classical}).
This demonstrates the importance of understanding \emph{how} the disturbance vanishes in the
weak limit.

To gain a better understanding of this problem, we find it useful to develop the theory
of both classical and quantum weak measurements.
The parallels between quantum
and non-ideal classical measurements have recently been highlighted\cite{DJ12B,JKM14},
see also Refs.~\cite{RSH91,BBKN13,DBN14}.
We start by introducing a notion of \emph{generalized weak measurements} within
an operational framework\footnote{By  \emph{operational} we mean that 
  the framework is expressed in terms of notions directly related to the experimental
  situation, e.g.~preparation procedures and outcome probabilities.}
which is general enough to encompass both classical and
quantum mechanics. 
When applied to a classical system, the (generalized) weak 
expectation value takes the form
\be
  \frac{\sum_{j,k}q_k\tilde{A}_{kj}p_j}{\sum_j q_jp_j}.\qquad \text{(classical)}
  \label{eq:intro-classical}
\ee
Where $\tilde{A}$ is a real matrix, $p_j$ is the probability for the system to be
prepared in state $j$, and $q_j$ is the probability for the postselection to
succeed given the system is in state $j$. In the quantum case the generalized weak 
expectation value takes the form
\be
  \re \frac{\langle\phi|\hat A|\psi\rangle}
    {\langle\phi|\psi\rangle},\qquad \text{(quantum)}
  \label{eq:intro-quant}
\ee
which is the standard AAV form\cite{AAV88}, \emph{except} that $\hat A$ is not 
Hermitian in general. Weak values of non-Hermitian operators have been
considered previously\cite{Wis02,LR05,GJM13,BC14}. We show that for any real matrix $\tilde A$ 
there is a weak measurements procedure that yields \eqref{eq:intro-classical} as 
its expectation
value by an explicit example. Similarly we show that any operator $\hat A$ can in
principle appear in \eqref{eq:intro-quant}.

While \eqref{eq:intro-classical} and \eqref{eq:intro-quant} look similar, the 
possibility of (quantum) interference in the denominator of \eqref{eq:intro-quant}
makes an important difference. In particular the classical weak measurement
only exhibits anomalous weak values when $\tilde A$ is not diagonal 
(which implies the the measurement process disturbs the system), while anomalous
weak values occur in the quantum case for \emph{any} non-trivial (i.e.~not proportional
to the identity) $\hat A$.

For the generalized weak measurements, the only constraint on the disturbance induced 
by the measurement is that it should vanish at vanishing interaction. To control the
disturbance we introduce two different constraints in the general framework. Both
constraints lead to the usual notion of a non-disturbing measurement when applied
to classical mechanics. For quantum mechanics the situation is more intriguing;
One of the constraints is impossible to satisfy, while the other one can be 
satisfied by \emph{all} measurements (strictly speaking you have to change the
measurement procedure slightly, but this change has no effect on the actual measurement
outcome).

Another way to control the disturbance is to introduce some quantitative measure of the
amount of disturbance. We introduce such a measure following Ref.~\cite{Ban01}, and we 
show under quite general assumptions that the measurements minimizing this quantity lead
to the usual AAV weak value. This results should be compared to the uniqueness theorem 
of Ref.~\cite{DJ12C}.

Disturbance in weak measurements has previously been analyzed through
Leggett-Garg inequalities\cite{RKM06,JKB06,WJ08}.
In particular it has been shown\cite{WJ08} that anomalous weak values imply
either that the measurement is disturbing or that \emph{macrorealism} fails.
In this paper we will take a purely operational point of view, and 
as a consequence assumptions
such as macrorealism will play no role.

The outline of the article is as follows: In Section \ref{sec:gen-formalism}
we introduce generalized weak measurements in a general operational formalism.
We then consider weak measurements in classical mechanics in Section \ref{sec:classical}.
Section \ref{sec:quant} forms the main part of these notes and deals with 
weak measurements in quantum mechanics.
We end in Section \ref{sec:discussion} with a discussion and outlook on some questions that
would be interesting to address in further work. 
Appendix \ref{sec:generalizations} addresses some ways to generalize the formalism,
while appendix \ref{sec:generalized-AAV} deals with the special case of 
von Neumann measurements. Finally
appendices \ref{sec:generalized-observable}
and \ref{sec:uniqueness-proof} contains some technical details.

\section{General formalism}
\label{sec:gen-formalism}
Here we will described generalized weak measurements in an operational 
framework which is independent of the details of the physical system under
consideration. Our framework can be seen as a variation of the
\emph{Generalized Probabilistic Theories}, see Refs.~\cite{BBLW12,Pfi12}
for recent expositions.

Let $\states$ be the set of preparation procedures of the system.
In classical mechanics an element of $\states$ would be a probability distribution
on the systems phase space, while for quantum systems the elements are density
matrices. For brevity we will often refer to the elements of $\states$ as states.
Our measurement apparatus will have a finite number of outcomes, and we will use
the index $m$ to denote a specific outcome. Given a state $s\in\states$, the 
probability of getting outcome $m$ is denoted $P^\lambda(m|s)$. The non-negative number
$\lambda$
quantifies the interaction strength between the system and the apparatus. The important 
point is that both the disturbance caused by the measurement apparatus and the information
extracted about the system should go to zero as $\lambda\to 0$. 

In order to define expectation values, we need to assign numerical values to the 
measurement outcomes.
We thus introduce a real number $A_m$ to each $m$, and, considering $A$ as a random variable, we define
the expectation value
\be
  \lexpval := \sum_m A_m P^\lambda(m|s).
\ee
The $A_m$ can be understood as \emph{contextual values} as introduced in Ref.~\cite{DAJ10}
(see also Ref.~\cite{DJ12B}).
In these notes we will only be interested in the \emph{weak limit} $\lambda\to 0$.
With no interaction, $\lambda = 0$, the probability $P^\lambda(m|s)$ is assumed to be independent of
the state $s$, and will be denote $P^0(m)$. We further assume that we have an 
asymptotic expansion  around $\lambda = 0$,
\be
  P^\lambda(m|s) = P^{0}(m)+\lambda\delta P(m|s) + O(\lambda^2).
\ee
For simplicity we assume that
\be
  \lexpvalz = \sum_m A_m P^0(m) = 0.
  \label{eq:A-constraint}
\ee
We can then define the following (non-postselected) weak limit
of the expectation value:
\be
  \wexpval := \lim_{\lambda\to 0}\lambda^{-1}\lexpval
  = \sum_m A_m\delta P(m|s).
\ee
Note that we have to amplify the signal by a factor $\lambda^{-1}$ to get
something non-trivial.
As discussed in Appendix \ref{sec:generalizations} there is no loss of generality
in assuming \eqref{eq:A-constraint}, and we will continue doing so in the following.
%In the remainder of these notes we will tacitly assume that Eq.~\eqref{eq:A-constraint}
%is satisfied. 

In order to discuss postselection we need to know the state of the system once it
leaves the measurement apparatus. The state after the measurement conditioned on a
given outcome is specified by the map
\be
  s \mapsto s' = M_m^\lambda(s).
\ee
Note that the map is non-trivial even for non-disturbing measurements, since the outcome
$m$ in general increases our knowledge about the system.
We denote the the of postselection procedures by $\dstates$. An element of $\dstates$
is a map\footnote{Note that $\dstates$ does not contain all maps $\states \to [0,1]$.}
$\sdual: \states \to [0,1]$ giving the probability that the postselection will
succeed on a given state,
\be
  \sdual(s) := P(\text{$\sdual$ will accept $s$}).
\ee
By only considering the experimental runs where a given postselection procedure succeeds,
we get the following expectation value (the product $P^\lambda(m|s)\sdual(M_m^\lambda(s))$ is
a joint probability, in more standard notation it might be written $P(m,\sdual|s)$)
\be
  \lexpvalp 
    := \frac{\sum_m A_m P^\lambda(m|s)\sdual(M_m^\lambda(s))}
      {\sum_m P^\lambda(m|s)\sdual(M_m^\lambda(s))}.
  \label{eq:lexpvalp-def}
\ee
In words $\lexpvalp$ is the conditional expectation value
of $A$ given a initial preparation $s$, and conditioned on the success of a
final postselection $\sdual$.
To take the weak limit of this we need to demand that 
\be
  M^{\lambda=0}_m(s) = s, \qquad \text{for all $m$, $s$,}
\ee
in accordance with our interpretation of $\lambda$ as interaction strength.
We can then define the \emph{generalized weak value} by
\be
  \wexpvalp 
  := \lim_{\lambda\to 0}\lambda^{-1}\lexpvalp
  = \wexpval + \sum_m A_m\frac{P^0(m)\delta\sdual(M_m(s))}{\sdual(s)}
  \label{eq:gen-weak},
\ee
where $\delta\sdual(M_m(s))$ is defined by the following small $\lambda$
expansion:
\be
  \sdual(M_m^\lambda(s)) = \sdual(s) + \lambda\delta\sdual(M_m(s)) + O(\lambda^2).
\ee
The RHS of \eqref{eq:gen-weak} is only defined when $\sdual(s)$ is non-zero, and this
will be tacitly assumed in the following.

We will  use that $\states$ and $\dstates$ are convex set. I.e.~if
$s$ and $s'$ are preparation procedures, then one can construct a combined procedure
by selecting procedure $s$ with probability $\alpha$ and $s'$ with probability $1-\alpha$.
This combined state is denoted $\alpha s+(1-\alpha)s'$.
By a similar construction $\dstates$ is also convex. From basic probability theory we get
the following relations:
\be
  P^\lambda(m|\alpha s+(1-\alpha)s') = \alpha P^\lambda(m|s) + (1-\alpha) P^\lambda(m|s'),
\ee
\be
  P^\lambda(m|\alpha s+(1-\alpha)s')M^\lambda_m(\alpha s+(1-\alpha)s') 
    = \alpha P^\lambda(m|s) M^\lambda_m(s) + (1-\alpha) P^\lambda(m|s') M^\lambda_m(s'),
\ee
and
\be
  \sdual(\alpha s+(1-\alpha)s') = \alpha \sdual(s) + (1-\alpha) \sdual(s'),\qquad
  (\alpha\sdual+(1-\alpha)\sdual')(s) = 
    \alpha \sdual(s)+(1-\alpha)\sdual'(s).
\ee
From these relations it follows that the function
\be
  G(s,\sdual) := \sdual(s)\, \wexpvalp
  \label{eq:G-def}
\ee
is bilinear with respect to convex combinations,
\be
  G(\alpha s+(1-\alpha)s',\sdual) = \alpha G(s,\sdual) + (1-\alpha) G(s',\sdual),
  \label{eq:G-convex1}
\ee
and
\be
  G(s,\alpha\sdual+(1-\alpha)\sdual') = \alpha G(s,\sdual) + (1-\alpha) G(s,\sdual').
  \label{eq:G-convex2}
\ee

\subsection{Disturbance}
\label{sec:gen-disturbance}
Without postselection we do not need to worry about how
the measurement apparatus affects the system, but,
as we have seen, the generalized weak expectation value \eqref{eq:gen-weak}
will depend on this disturbance.
In order to associate an unique postselected expectation value with a
given ordinary observable we thus have to constrain the disturbance. Here
we formulate two simple condition within the general operational framework. Later we will
see that both of these have the desired effect on classical measurements, but that
the situation is not so simple for quantum mechanics.

Morally, we want to say that the measurement apparatus does not change the ontic
state of the system, but since our operational framework lack the notion of an
ontic state, we cannot
express this directly. Instead we can assume that there exists a subset of the
states $\bstates\subset\states$ such that every state $s$ can be written as
a convex combination of states in $\bstates$,
\be
  s = \sum_j p_j s_j, \qquad s_j \in\bstates,\qquad\sum_j p_j = 1.
  \label{eq:s-expansion}
\ee
We will then say that a measurement procedure is 
\emph{non-disturbing in the strong sense} if
\be
  M_m^\lambda(s) = s + O(\lambda^2)\qquad\text{for all $s\in\bstates, m$}.
  \label{eq:strongly-non-disturbing}
\ee
Assume that we have an expansion of $s$ as in \eqref{eq:s-expansion}. Given
a $\sdual$ we can then define a new state by
\be
  \scombined := (\sdual(s))^{-1}\sum_j p_j\sdual(s_j) s_j.
\ee
If now \eqref{eq:strongly-non-disturbing} holds, we find
\be
  \wexpvalp = \wexpvalc 
    := \frac 1 {\sum_j p_j \sdual(s_j)}
      \sum_j p_j \sdual(s_j) \wexpvalj.
\ee
Thus, if a measurement is non-disturbing in the strong sense, then the postselected
weak value is equal to the non-postselected weak value in the combined ensemble
$\scombined$.

Another possibility is to say that the state we obtain 
\emph{if we ignore the measurement outcome $m$}, i.e.
\be
  M^\lambda_?(s) := \sum_m P^\lambda(m|s)M^\lambda_m(s),
\ee
is just $s$ to first order in $\lambda$. We will thus call a measurement procedure
such that
\be
  M^\lambda_?(s) = s + O(\lambda^2)
  \label{eq:weakly-non-disturbing}
\ee
\emph{non-disturbing in the weak sense}. This definition is adopted in
Ref.~\cite{Hof10,BC14}. We note that \eqref{eq:strongly-non-disturbing}
indeed implies \eqref{eq:weakly-non-disturbing} in accordance with the naming.

\section{Classical mechanics}
\label{sec:classical}
In order to clarify the ideas of the previous section, and to provide a background
to understand quantum weak measurements, let us consider the situation in classical mechanics.
A model of weak measurements with disturbance on a classical system was recently 
given in Ref.~\cite{FC14}. That model does, however, not strictly fall within our
framework, since the dependence of the disturbance on the interaction strength is different.
Models of weak measurements on classical \emph{fields} have also been 
considered\cite{RSH91,BBKN13,DBN14}. A conceptual difference between the models we will
consider and the field models is that for the field models the measurement disturbance
is deterministic, while we will only consider stochastic disturbance.

For simplicity we will consider systems with a finite number
of ontic states (i.e.~the `phase space' of the system consists of a finite number of points),
and we will denote these $s_j$.
The preparation procedures are then specified by probability distributions on the
ontic states, that is
\be
  \states = \left\{\sum_j p_j s_j \,\middle|\, \sum_j p_j = 1, p_j\geq 0\right\},
\ee
where $p_j$ is the probability of preparing the system in state $j$.
Defining dual states by
\be
  \sdual_j(s_k) := \delta_{jk},
\ee
we can also expand $\sdual$ as
\be
  \sdual = \sum_j q_j \sdual_j,\qquad q_j := \sdual(s_j).
\ee
Using the bilinearity of the $G(s,\sdual)$ function (Eq. \eqref{eq:G-def}), we
find
\be
  G(s,\sdual) = \sum_{j,k}q_k\tilde{A}_{kj}p_j,
\ee
with the real matrix $\tilde A$ defined by
\be
  \tilde A_{kj} := \lim_{\lambda\to 0}\lambda^{-1}\sum_m A_m 
        P^\lambda(m|s_j)\sdual_k(M_m^\lambda(s_j)).
\ee
It follows immediately that the generalized weak value is
\be
  \wexpvalp = \frac{1}{\sdual(s)}\sum_{j,k}q_k\tilde{A}_{kj}p_j.
  \label{eq:classical-weak}
\ee

A natural question is whether all real matrices $\tilde A$ can appear
in \eqref{eq:classical-weak}? The answer is positive, as can be seen by the
following simple construction. Let the real matrix $\tilde A$ be given.
 We consider a measurement with
two outcomes, denoted by $m=\pm$. Take the probability to get a given outcome
to be (note that $\lambda$ has to be sufficiently small for the model to make sense)
\be
  P^\lambda(m=\pm|s_j) = \frac 1 2 \pm \frac {\lambda} 2\sum_k \tilde{A}_{kj},
\ee
and the post-measurement state to be
\be
  M_{\pm}^\lambda(s_j) = (1-2\lambda\sum_{k\neq j}[\pm\tilde{A}_{kj}]_+)s_j
    +2\lambda\sum_{k\neq j}[\pm\tilde{A}_{kj}]_+ s_k.
\ee
In the last equation $[\cdot]_+$ denotes the positive part, as defined by
\be
  [x]_+ := \max\{x,0\}.
  \label{eq:positive-part}
\ee
A calculation now shows that \eqref{eq:classical-weak} is indeed satisfied.
We conclude that the space of generalized weak measurements on a classical
system with $d$ states is in one-to-one correspondence with the space of real $d\times d$ 
matrices\footnote{Here we identity measurement procedures where the weak values
  are identical, i.e.~where the maps $(s,\sdual)\mapsto \wexpvalp$ agree.}.

Before we turn to quantum mechanics let us note  the following  result:
 if a classical weak measurement is non-disturbing in the weak sense
if and only if it is non-disturbing in the strong sense. One direction has already
been shown to hold in general. To see the other direction we assume that the
measurement is non-disturbing in the weak sense. We take $\bstates$ to be the
set of ontological states. By assumption we have 
\be
  M^\lambda_?(s) := \sum_m P^\lambda(m|s)M_m^\lambda(s) = s + O(\lambda^2)
  \label{eq:classical-M?}
\ee
for all states $s\in\states$. Using that every state can 
\emph{uniquely}\footnote{This uniqueness fails in the quantum mechanical case.} be 
written as
\be
  s = \sum_j p_j s_j,\qquad s_j\in \bstates
\ee
it is now easy to check that \eqref{eq:classical-M?}
can only hold for ontic states $s\in\bstates$ if we have
\be
  M_m^\lambda(s) = s + O(\lambda^2)\qquad\text{for all $s\in\bstates, m$}.
\ee
Going back to \eqref{eq:classical-weak} we see that for non-disturbing 
classical weak measurements $\tilde A$ will be diagonal (the converse
is however not true in general).

For classical mechanics we thus have the following simple picture: If a generalized
weak measurement is non-disturbing in the usual sense that it does not change the
ontic state of the system, then it will be described by a diagonal matrix
$\tilde A_{kj}$ (furthermore is easy to see that all diagonal matrices appear this
way). By the above result it is actually sufficient to assume that the measurement is 
non-disturbing in the weak sense. If one does not put any constraints on the disturbance,
then the measurement is described by a general real matrix $\tilde A_{kj}$.

\section{Quantum mechanics}
\label{sec:quant}
Having discussed the simpler classical case, we go on the main topic of the paper, 
namely weak measurements in quantum mechanics. We take it as an axiom of quantum
mechanics that  the space of
preparation procedures, $\states$, is identified with the set of density matrices
(positive operators of 
trace one) on some Hilbert space $\hilb$,
\be
  \states = \left\{s\in\ndo(\hilb)\middle| s=s^\dagger,\; s\geq 0,\; \tr[s] = 1\right\}.
\ee
In the remainder of the article we
will keep the finite dimensional system space $\hilb$ fixed. For the set of postselection
conditions the most general choice is the \emph{effects} 
on $\hilb$. We will thus take $\sdual$ to be a positive operator with 
eigenvalues $\leq 1$, 
\be
  \dstates = \left\{\sdual\in\ndo(\hilb)\middle| \sdual=\sdual^\dagger,\; 0\leq\sdual\leq 1 \right\}.
\ee
The probability for a system in state $s$ to be postselected
is then
\be
  \sdual(s) := \tr [\sdual s].
\ee
In particular, having no postselection (i.e.~accepting all runs of the experiment)
is represented by setting $\sdual = \idop$.

Before we perform an explicit calculation of $\wexpvalp$,
let us anticipate the result using a more heuristic
argument. We recall that the function $G$ satisfies 
\be
  G(\alpha s+(1-\alpha)s',\sdual) = \alpha G(s,\sdual) + (1-\alpha) G(s',\sdual),
\ee
and
\be
  G(s,\alpha\sdual+(1-\alpha)\sdual') = \alpha G(s,\sdual) + (1-\alpha) G(s,\sdual').
\ee
The simplest non-trivial family of real functions with this
property is $\re\tr[\sdual\hat A s]$, where $\hat A$ is a
(not necessarily Hermitian) operator on $\hilb$. One could also
imagine having terms of the form $\re\tr[\sdual\hat A s \hat B]$, but because
we only expand to first order in $\lambda$ we will not see this more general type
of term, however see Appendix \ref{sec:generalizations}.
We thus claim that the post-selected weak value must take the form
\be
  \wexpvalp = \frac{\re\tr[\sdual\hat A s]}{\tr[\sdual s]}.
  \label{eq:quant-weak}
\ee
Note that this expression has both the 
real and imaginary part of the usual weak value as special cases.
Indeed, if we set $s = |\psi\rangle\langle\psi|$, $\sdual = |\phi\rangle\langle\phi|$ 
and $\hat A = \hat O$, where $\hat O$ is Hermitian, we
recover the real part of the usual AAV expression\cite{AAV88}
\be
  \wexpvalp =  \re \frac{\langle\phi|\hat O|\psi\rangle}
    {\langle\phi|\psi\rangle}.\qquad (\hat A = \hat O)
\ee
On the other hand, setting $\hat A = -i\hat O$, we obtain the imaginary part
\be
  \wexpvalp =  \im \frac{\langle\phi|\hat O|\psi\rangle}
    {\langle\phi|\psi\rangle}.\qquad (\hat A = -i\hat O)
\ee
%except that we don't require $\hat A$ to be Hermitian. 
We will call $\hat A$ a \emph{generalized observable}.

Some operators give the same expectation values when
plugged in to \eqref{eq:quant-weak}. To be precise one should thus define a generalized
observable to be a element of
\be
  \ndo(\hilb)/\sim,
  \label{eq:fancy-gen-observable}
\ee
where $\hat A \sim \hat A'$ iff $\hat A - \hat A'$ is a purely imaginary multiple of the
identity. See Appendix \ref{sec:generalized-observable} for further details.

Let us now verify \eqref{eq:quant-weak} by a more careful calculation.
The most general measurement on a quantum system can be described by a \emph{quantum instrument}
\cite{Da76}. For our purposes it will be convenient to express the instrument in terms
of Kraus operators. For each measurement outcome $m$ we thus have a family of operators
$\khat^\lambda_{m,n}$ on $\hilb$ such that
\be
  \sum_{m,n}(\khat^\lambda_{m,n})^\dagger \khat^\lambda_{m,n} = \idop.
  \label{eq:K-norm}
\ee
The probability of obtaining outcome $m$ is
\be
  P^\lambda(m|s) := \sum_n \tr[s(\khat^\lambda_{m,n})^\dagger \khat^\lambda_{m,n}],
  \label{eq:quant-P}
\ee
and the post-measurement state is
\be
  M_m^\lambda(s) := \frac{\sum_n \khat^\lambda_{m,n}s(\khat^\lambda_{m,n})^\dagger}{P^\lambda(m|s)}.
  \label{eq:quant-M}
\ee
We assume that the Kraus operators have an expansion in $\lambda$, 
\be
  \hat K^\lambda_{m,n} = K^0_{m,n}+\lambda\delta \khat_{m,n} 
    + \frac 1 2 \lambda^2 \delta^2 \khat_{m,n} + O(\lambda^3).
\ee
The basic assumption that $M^{\lambda=0}_m(s) = s$ is then equivalent to 
\be
  K^0_{m,n} \propto \idop,\qquad \text{for all $m,n$.}
\ee
It is clear from \eqref{eq:quant-P} and \eqref{eq:quant-M}
that the physics is invariant under a change of phase of the $\hat K_{m,n}^\lambda$
operators.
We will thus assume that $K^0_{m,n}$ is real and positive (for all $m,n$).
Plugging \eqref{eq:quant-P} and \eqref{eq:quant-M} into \eqref{eq:gen-weak} 
we obtain \eqref{eq:quant-weak} with $\hat A$ explicitly given by 
\be
  \hat A := 2\sum_m A_m \delta \bar{K}_m, 
  \label{eq:A-hat}
\ee
and where we define the averaged $\delta \khat$ by
\be
  \delta \bar{K}_m % = \delta \bar{K}_m^H + \delta \bar{K}_m^{aH} 
  := \sum_n K^0_{m,n}\delta \khat_{m,n}.
\ee

Similarly to the classical case, we can show that any generalized observable 
$\hat A$ is realized by a measurement scheme. To show this we consider 
the following explicit model, which has been previously discussed in 
Ref.~\cite{GJM13}:
Let $\hat A \in \ndo(\hilb)$ be
given, and let $\hilba$ be a two dimensional Hilbert space with orthonormal basis
$|\pm\rangle$. On $\hilb\otimes\hilba$ we define the operator
\be
  2\hat H := i\hat A^R\otimes |-\rangle\langle +|
    -i\hat A^R\otimes |+\rangle\langle -|
    +\hat A^I\otimes |+\rangle\langle +|
    -\hat A^I\otimes |-\rangle\langle -|.
  \label{eq:qubit-model-H}
\ee
The model is then defined by setting (we omit the $n$ index on $\khat$, since
it is trivial)
\be
  \khat_{\pm}^\lambda = \frac{1}{\sqrt 2}\tr_{\text{aux}}
    [e^{i\lambda H}(|+\rangle+|-\rangle)\langle \pm|]
    = \frac{1}{\sqrt{2}}\pm\frac{\lambda}{2\sqrt{2}}\hat A+O(\lambda^2),
  \label{eq:qubit-model-K}
\ee
\be
  A_{\pm} = \pm 1,
\ee
and \eqref{eq:quant-weak} is verified. Physically the model can be understood
as letting the system $\hilb$ interact weakly with an auxiliary qubit, and then
performing a projective measurement on the qubit. It is easy to show that this model
is non-disturbing in the weak sense, for all operators $\hat A$.

Let us rewrite the expression for the (generalized) quantum weak value
in a way that makes comparison with the classical case easier.
We will focus on pure states, so we set
$s = |\psi\rangle\langle\psi|$ and $\sdual = |\phi\rangle\langle\phi|$.
Choose an orthonormal basis $|j\rangle$ for $\hilb$, and define
\be
  u_j = \langle j|\psi\rangle,\quad v_j = \langle j|\phi\rangle,\quad
  \hat{A}_{kj} := \langle k|\hat A|j\rangle.
\ee
The weak value is then given by
\be
  \wexpvalp = \re\frac{\sum_{jk}v^*_k \hat{A}_{kj} u_j}
                                    {\sum_j v^*_j u_j}. \qquad \text{(quantum)}
  \label{eq:quant-a-la-classical}
\ee
On the other hand, the classical weak value is given by (Eq. \eqref{eq:classical-weak})
\be
  \wexpvalp = \frac{\sum_{jk}q_k \tilde{A}_{kj} p_j}
                                    {\sum_j q_j p_j}. \qquad \text{(classical)}
\ee
The two expressions look very similar, but it is important to keep in mind that
$p_j$ and $q_j$ are (positive) probabilities, while $u_j$ and $v_j$ are (complex)
amplitudes. This makes an important difference. Let us say that a measurement allows
for \emph{anomalous weak values} if one can make $\wexpvalp$ 
arbitrarily large by choosing $s$ and $\sdual$ appropriately. In the classical 
case we see that this is possible iff $\tilde A_{jk}$ is not diagonal (anomalous
weak values in classical systems are also discussed in Ref.~\cite{FC14}).
In the quantum case, however, we can get anomalous weak values for any non-trivial
(i.e. not proportional to the identity) $\hat A_{jk}$ due to the possibility of
destructive interference in the denominator of \eqref{eq:quant-a-la-classical}.

In this section we avoid discussing the details of the 
measurement apparatus. Since the concept of weak measurement is often presented in 
the context of von Neumann measurements, we
consider this case in detail in Appendix \ref{sec:generalized-AAV}.

\subsection{(Non-)Disturbance in the weak and strong sense}
Let us first show that a non-trivial weak measurement cannot be
non-disturbing in the strong sense. 
%This can be understood as an weak analogue
%of the ``no information without disturbance'' theorem for strong measurements 
%(see Theorem 2 of Ref. \onlinecite{Bus09}).
In order that every state can be written as
\be
  s = \sum_j p_j s_j, \qquad s_j \in\bstates,
\ee
it is well known that $\bstates$ must contain all pure states\footnote{Note that
  $\bstates$ must then be (uncountably) infinite, but that we will still
  only need to consider finite sums of states from $\bstates$}.
To first order in $\lambda$, $M_m^\lambda$ sends pure states
to pure state:
\be
  M_m^\lambda(|\psi\rangle\langle\psi|)
    = |\psi'\rangle\langle\psi'| + O(\lambda^2)
\ee
with
\be
  |\psi'\rangle = \left(1+\frac{\lambda}{P^0(m)}[\delta\bar K_m
                         -\langle\psi|\delta\bar K_m|\psi\rangle\idop]\right)
                      |\psi\rangle.
  \label{eq:psi'}
\ee
The only way that $|\psi'\rangle$ can be in the same
ray as $|\psi\rangle$ for all $m$ and $\psi$ is if all $\delta\bar K_m$
are proportional to the identity. But then we also have $\hat A \propto \idop$ 
and $\wexpvalp$ becomes a trivial constant independent of $s$ and $\sdual$.

The situation for the weak condition of Section \ref{sec:gen-disturbance} is quite
different. We fist note that
\be
  M^\lambda_?(s) := \sum_m P^\lambda(m|s) M_m^\lambda(s) 
    = s +\lambda i[\hat D,s] + O(\lambda^2),
  \label{eq:quant-M?}
\ee
where 
\be
  \hat D := % \sum_m\delta\bar{K}_m^{I} = 
  -i\sum_m\delta\bar{K}_m.
  \label{eq:D-def}
\ee
Here we have used that from \eqref{eq:K-norm} it follows that 
$\hat D$ is Hermitian. Since \eqref{eq:quant-M?} is a unitary
transformation to order $\lambda$, we can eliminate the disturbance by performing
the inverse unitary after $M$. Moreover, this compensating transformation does
change the generalized observable $\hat A$. In more detail, the replacement
\be
  \khat_{m,n}^\lambda \to e^{-i\lambda\hat D}\khat_{m,n}^\lambda
  \label{eq:khat-unitary}
\ee
ensures that $M_?^\lambda(s) = s + O(\lambda^2)$ and using \eqref{eq:A-constraint}
one can check that it leaves $\hat A$ invariant.

From \eqref{eq:quant-M?} it follows that the overall probability for successful
postselection is
\be
  \sdual(M^\lambda_?(s)) 
    = \tr[\sdual s]\left(1-2\lambda\im\frac{\tr[\sdual\hat D s]}{\tr[\sdual s]}\right)
      + O(\lambda^2).
\ee
Thus, the relative change (due to the intermediate measurement) of the postselection
probability is  proportional to the imaginary part of the AAV weak value of
$\hat D$ (see also \cite{DJ12A}). Note that in the general setting we are considering 
there need not
to be any connection between $\hat D$ and $\hat A$, but for von Neumann
measurements one has $\hat D \propto \hat A$, see Ref. 
\cite{DJ12A} and Appendix \ref{sec:generalized-AAV}.

We see that disturbance in quantum mechanics behaves quite different from 
classical mechanics.
One the one hand a quantum mechanical measurement cannot be non-disturbing
in the strong sense (except in the trivial case), whereas this is usually implicitly 
assumed for classical measurements.
On the other hand being non-disturbing in the weak sense is rather restrictive in the classical 
setting (since it implies being non-disturbing in the strong sense), while it does not restrict
the class quantum mechanical measurements at all (in the sense that the generalized 
observable $\hat A$ is unconstrained).

\subsection{Minimal disturbance and uniqueness of the weak value}
\label{sec:uniqueness}

By setting $\sdual = \idop$ in \eqref{eq:quant-weak} we obtain the expectation value without
postselection,
\be
  \wexpval = \tr[s\hat A^R] =: \qxpval{\hat A^R}_s.
  \label{eq:quant-no-post}
\ee
Here $\hat A^R$ denotes the Hermitian part of $\hat A$, i.e.
\be
  \hat A = \hat A^R + i\hat A^I, \qquad (\hat A^R)^\dagger = \hat A^R, \qquad
    (\hat A^I)^\dagger = \hat A^I.
\ee
We have just seen that the conditions of non-disturbance discussed in 
Section \ref{sec:gen-disturbance} are not useful in restricting the
allowed generalized observable. This means that, given an ordinary observable
$\hat O$, it is not given which generalized observable $\hat A$ (satisfying 
$\hat A^R = \hat O$) we should associate with it. This is in contrast to
the classical case, where either of the conditions of non-disturbance selects
a unique $\tilde A$ (namely the diagonal one) for a given observable.
For an extended discussion of the uniqueness of the weak value see Ref.~\cite{DJ12C}
and references therein.

Instead of requiring the measurement to be non-disturbing, one can look for
for a way to quantify the amount of disturbance, and then demand  this
quantity to be minimal. In Refs.~\cite{DAJ10,DJ12C} it is shown that one recovers the
AAV weak value if one requires that the Kraus operators are positive and Hermitian
(this is taken as the definition of a minimally disturbing measurement in 
Ref.~\cite{WM09}). Note that the assumptions of Refs.~\cite{DAJ10,DJ12C} are
somewhat different from ours.\footnote{For instance it is assumed in
  Refs.~\cite{DAJ10,DJ12C} that the effect operators 
  (i.e. $\sum_n (\khat_{m,n}^\lambda)^\dagger \khat_{m,n}^\lambda$ in our notation) all commute
  with the observable.}

Here we want to highlight a numerical quantity measuring disturbance\cite{Ban01}
which is minimized, and show how it appears from a operational point of view. 
Note that, in contrast to the  various error-disturbance relations
discussed recently (see e.g.~\cite{Oza03,BH08,Bra13,BLW14,BHOW14,Ips13}),
here we are interested in the disturbance of the system as such,
rather than one of its observables.
In fact, there is no good candidate for the observable in the case we are considering (in 
particular $\hat A^R$ would be a bad choice, since then the disturbance would simply
be zero for a large class of measurement procedures).

For a system prepared in a pure state, a natural way to measure 
the disturbance is by the \emph{survival probability} (alternatively,
the quantum fidelity between the initial and final state)
\be
  F^\lambda(\psi) := \tr \left[M_?^\lambda(|\psi\rangle\langle\psi|) |\psi\rangle\langle\psi|\right].
\ee
This is simply the probability that the system was not kicked into an orthogonal state
by the measurement process. 

Expanding in $\lambda$ we find that
\be
  F^\lambda(\psi)
    = 1-\lambda^2\sum_{m,n}\left(\langle \psi | \delta \khat_{m,n}^\dagger \delta \khat_{m,n} | \psi\rangle
    -|\langle \psi| \delta \khat_{m,n} | \psi\rangle |^2\right) + O(\lambda^3).
\ee
Here we have simplified the expression using the relation
\be
  \sum_{m,n}(\delta \khat_{m,n})^\dagger \delta \khat_{m,n}
  + \frac 1 2 \sum_{m,n} K_{m,n}^0 [\delta^2  \khat_{m,n}^\dagger +  \delta^2  \khat_{m,n}] = 0,
\ee
which follows from \eqref{eq:K-norm}. Note that the leading order term of $F^\lambda$ only
depends on the first order terms of $\khat_{m,n}^\lambda$. To get a state independent
number we now average over $\psi$ with respect to the Haar measure\cite{Ban00,Ban01}.
We use the integral
\be
  \int\ud\psi\, \langle\psi|\hat B |\psi\rangle\langle\psi| \hat C |\psi\rangle
    = \frac{1}{d(d+1)} (\tr[\hat B\hat C]+\tr[\hat B]\tr[\hat C]),
\ee
and find
\be
  \bar F^\lambda := \int\ud\psi\, F^\lambda(\psi) = 1-\frac{\lambda^2}{d(d+1)}\distmeas
    + O(\lambda^3),
\ee
with 
\be
  \distmeas := \sum_{m,n}
    \left(d\tr[\delta\khat_{m,n}^\dagger \delta\khat_{m,n}]-|\tr[\delta\khat_{m,n}]|^2\right).
\ee
We will take $\distmeas$ as our measure of disturbance. Note that $\distmeas$ can be understood
as a weak limit of $\bar F^\lambda$,
\be
  \distmeas = d(d+1)\lim_{\lambda\to 0}\lambda^{-2}(1-\bar F^\lambda).
\ee

It is convenient to write
\be
  \distmeas = \sum_{m,n}\distmeasf(\delta\khat_{m,n}^R)+\distmeasf(\delta\khat_{m,n}^I)
  \label{eq:distmeas-split},
\ee
where
\be
  \distmeasf(\hat B) := d\tr[\hat B^2]-(\tr[\hat B])^2, \qquad \text{for Hermitian $\hat B$}.
\ee
The function $\distmeasf(\hat B)$ is non-negative, and vanishes iff $\hat B$ is
proportional to the identity. It follows immediately that $\distmeas$ is 
strictly positive for all non-trivial measurements. We can now show
the following (the proof and exact statement is in Appendix \ref{sec:uniqueness-proof}):
Fix the number of measurement outcomes
and an observable $\hat O$. Bound (or fix) the values $A_m$. Among the generalized weak measurements
with $\hat A^R = \hat O$ those which minimize $\distmeas$ have $\hat A \sim \hat O$.
More loosely, the minimally disturbing generalized weak measurements yield the AAV
weak value.

As an explicit example, let us mention that for the model define by 
\eqref{eq:qubit-model-H} and \eqref{eq:qubit-model-K}, we find 
\be
 \distmeas = \frac{1}{4}\left(\distmeasf(\hat A^R)+\distmeasf(\hat A^I)\right).
\ee
Here we see explicitly that the disturbance is minimal exactly when $\hat A^I \propto \idop$.
In Appendix \ref{sec:generalized-AAV} we calculate $\distmeas$ for von
Neumann like models.

\section{Discussion}
\label{sec:discussion}
Let us outline some different attitudes one can take towards weak values
in light of the above remarks: 
\textbf{(a)} Generalized weak measurements that are
non-disturbing in the weak sense should be considered non-disturbing. By non-disturbing (without the
weak or strong qualifier) we mean that the disturbance is sufficiently weak that it \emph{does
not affect the weak value which is the result of the measurement.}
\textbf{(b)} All (non-trivial) generalized weak measurements should be considered disturbing.
The measurements of a given observable that are least disturbing yield the AAV weak
value.
\textbf{(c)} There are some generalized weak measurements that are non-disturbing,
and these always yield the AAV weak value.

Consider a weak measurement procedure which is non-disturbing in the weak
sense. Without postselection it will measure some ordinary observable $\hat O$.
If we consider the measurement to be non-disturbing, as postulated in option \textbf{(a)}, 
the intermediate measurement should not interfere with
postselection. Thus the experiment with postselection should still be a 
measurement of $\hat O$, just in a different ensemble (namely the one defined both by the
preparation and postselection). But $\wexpvalp$
also depends on $\hat A^I$, which is not determined by $\hat O$. In other words, 
the weak value depends on \emph{how} we measure $\hat O$, even though the measurement
is non-disturbing.
It seems that to understand option \textbf{(a)}, one is faced with the task of 
making sense of this additional  %\cite{KS67,Spe05}
dependence in the weak value. The relation between contextuality and weak values was
recently discussed in Ref.~\cite{Pus14}.

If weak measurements  disturb the system, then it is difficult to understand why the 
weak value should be considered the expectation value of an observable in 
the postselected ensemble.
We have seen that if one allows for (weak) disturbance in a classical setting, one
does not get the `right' answer when turning on postselection.
The main question arising from position \textbf{(b)} then seems to be: What is the 
fundamental interpretation of the weak value, other than the result of a specific
measurement procedure? Of course, it is possible that there is no such interpretation.

Option \textbf{(c)} is attractive because it allows for a straight forward 
interpretation of the weak value as the
expectation value of some observable between preparation and postselection.
The measure of disturbance $\distmeas$  lends some support
to this position in that, when it is minimal, the measurement yields the AAV weak value.
On the other hand, the minimum of $\distmeas$  \emph{cannot} be zero (unless the measured 
observable is a trivial constant), even in the original AAV setup (see also 
Eq. \eqref{eq:disturbance-bound}). It is possible that the exists ways of quantifying the 
disturbance such that option \textbf{(c)} is realized, but the author is not aware of any.

\begin{acknowledgments}
Josh Combes is thanked for useful discussions and for many helpful comments on a
draft of the present manuscript. Chris Ferrie, Aharon Brodutch and Alessandro Romito
are also thanked for useful discussion.
Financial support from 
the ERC-Advanced grant 291092 ``Exploring the Quantum Universe''
is acknowledged. This research was  supported in part by Perimeter Institute 
for Theoretical
Physics. Research at Perimeter Institute is supported by the Government of
Canada through Industry Canada and by the Province of Ontario through
the Ministry of Economic Development \& Innovation.

\end{acknowledgments}

\appendix

\section{Some further generalizations}
\label{sec:generalizations}
Here we discuss two generalizations of the framework considered in the main part
of the article. We will focus on the quantum case.
First, let us consider the constraint Eq. \eqref{eq:A-constraint},
\be
  \lexpvalz = \sum_m A_m P^0(m) = 0.
  \label{eq:A-constraint-again}
\ee
If we drop this constraint the conditional expectation value
\eqref{eq:lexpvalp-def} becomes
\be
  \lexpvalp = \lexpvalz + \lambda\frac{\re\tr[\sdual\hat A' s]}{\tr[\sdual s]} + O(\lambda^2),
  \label{eq:exp-val-no-constraint}
\ee
where $\hat A'$ contains an additional contribution proportional to $\lexpvalz$,
\be
  \hat A' := \hat A -i2\lexpvalz\hat D.
\ee
Here $\hat A$ is defined by Eq. \eqref{eq:A-hat}, while $\hat D$ is defined by
Eq. \eqref{eq:D-def}. It is now natural to define the generalized weak value to
be the coefficient of $\lambda$ in Eq. \eqref{eq:exp-val-no-constraint},
\be
  \wexpvalpp := \lim_{\lambda\to 0}\lambda^{-1}(\lexpvalp-\lexpvalz) =
    \frac{\re\tr[\sdual\hat A' s]}{\tr[\sdual s]}.
  \label{eq:quant-weak-no-constraint}
\ee
Let us now note that the shift
\be
  A_m \to A_m-\lexpvalz
\ee
leaves \eqref{eq:quant-weak-no-constraint} invariant while ensuring 
that \eqref{eq:A-constraint-again}
is satisfied. We thus conclude that there is no loss of generality in restricting to the
case where \eqref{eq:A-constraint-again} holds.

%With the replacement $\hat A\to \hat A'$ most of the results of the paper go 
%through. However, the discussion around \eqref{eq:khat-unitary} becomes a little more
%complicated. Let us also note that if the measurement is non-disturbing in the weak sense
%then $\hat D\propto \idop$, and hence $\hat A\sim \hat A'$. For weakly non-disturbing
%measurements one can thus drop the condition \eqref{eq:A-constraint-again} without
%changing anything.

A more substantial generalization comes about by reconsidering the asymptotic expansion
of the Kraus operators. In Section \ref{sec:quant} we assumed the $G(s,\sdual)$ function 
to take the form $G(s,\sdual) = \tr[\sdual\hat A s]$, however, the most general bilinear
real function takes the form
\be
  G(s,\sdual) = \sum_j \varepsilon_j\tr[\sdual \hat A_j s \hat A_j^\dagger],
  \label{eq:generalized-G}
\ee
where $\hat A_j$ is some set of (non-Hermitian) operators on $\hilb$ and $\varepsilon_j = \pm 1$.
This follows from the polarization identity
\be
  \re\tr[\sdual \hat A s \hat B]
  = \frac 1 4\left(\tr[\sdual (\hat A+\hat B^\dagger) s (\hat A+\hat B^\dagger)^\dagger]
    - \tr[\sdual (\hat A-\hat B^\dagger) s (\hat A-\hat B^\dagger)^\dagger]\right).
\ee
Terms of this more general form are obtained if some of the Kraus operators 
behave as 
\be
  \khat^\lambda_{m,n} = \lambda^{1/2}\hat L_{m,n} + O(\lambda^{3/2})
  \label{eq:Kraus-non-analytic}
\ee
in the weak limit. Note that \eqref{eq:Kraus-non-analytic} is compatible with $M_m^\lambda(s)$
and $P^\lambda(m|s)$ having expansions in integer powers of $\lambda$.
However, for indirect measurements where the Hamiltonian is an analytical function
of $\lambda$ (i.e.~von Neumann measurements or the qubit scheme discussed in 
Section \ref{sec:quant}) the Kraus operators will also be analytical in $\lambda$.

From \eqref{eq:generalized-G} it follows that
\be
  \wexpvalp = \frac{\sum_j \varepsilon_j\tr[\sdual \hat A_j s \hat A_j^\dagger]}{\tr[\sdual s]}.
  \label{eq:most-gen-weak}
\ee
A particular example of this is the so-called \emph{null weak values}\cite{ZRG12} where
\be
  \wexpvalp = \frac{\tr[\hat O s]}{\tr[\sdual s]},
\ee
for some Hermitian $\hat O$. The most general form \eqref{eq:most-gen-weak} can be obtained
by considering a measurement with two outcomes $m=\pm$. Indeed, setting $A_\pm = \pm 1$ and
(here $[\cdot]_+$ is defined by \eqref{eq:positive-part})
\be
  P^\lambda(m=\pm|s)M^\lambda_{\pm}(s) = \frac 1 2 s 
    +\lambda\sum_j\left([\pm\varepsilon_j]_+\hat A_j s \hat A_j^\dagger
      -\frac 1 4 \hat A_j^\dagger \hat A_j s -\frac 1 4 s \hat A_j^\dagger \hat A_j \right)
      + O(\lambda^2)
  \label{eq:most-gen-model}
\ee
one recovers \eqref{eq:most-gen-weak}. We leave the extension of the model \eqref{eq:most-gen-model}
to finite $\lambda$ to further work.

%The generalized expression \eqref{eq:most-gen-weak} allow us to understand the classical
%weak value as a special case of the quantum expression. 
Allowing Kraus operators of the form \eqref{eq:Kraus-non-analytic} we can embed the
classical model of weak measurements in the quantum model.
To see this, 
let us choose some basis $|j\rangle$ for the system
Hilbert space, and take $s$ and $\sdual$ to be diagonal,
\be
  s = \sum_j p_j|j\rangle\langle j|,\qquad
  \sdual = \sum_j q_j|j\rangle\langle j|.
\ee
With 
\be
  \hat A_{kj} := \sqrt{|\tilde A_{kj}|}|k\rangle\langle j|
\ee
we then find 
\be
  \frac{\sum_{j,k} \sgn(\tilde A_{kj})
    \tr[\sdual \hat A_{kj}
         s \hat A_{kj}^\dagger]}{\tr[\sdual s]}
  = \frac{\sum_{jk}q_k \tilde{A}_{kj} p_j}
                                    {\sum_j q_j p_j},
\ee
which is just the classical weak value \eqref{eq:classical-weak}.

Let us finally note that having Kraus operators with expansions of the form
\eqref{eq:Kraus-non-analytic} (with $\hat L_{m,n}$ not proportional to the identity)
implies that the measurement cannot be non-disturbing in the weak sense.

%In other words, for weakly non-disturbing measurements there is no loss of generality
%in restricting to the case considered in the main part of the article.

\section{`Gauge invariance' of generalized observables}
\label{sec:generalized-observable}
Given two generalized observables $\hat A, \hat A'$ we want to know whether they
give rise to the same expectation values, i.e. whether it holds that
\be
  \frac{\re\tr[\sdual\hat A s]}{\tr[\sdual s]} =
  \frac{\re\tr[\sdual\hat A' s]}{\tr[\sdual s]},\qquad 
  \text{for all $s \in \states,\sdual\in\dstates$ such that $\tr[\sdual s] \neq 0$}.
\ee
This is clearly equivalent to finding the operators $\hat B$ that satisfy
\be
  \re\tr[s\sdual\hat B] = 0,\qquad 
  \text{for all $s \in \states,\sdual\in\dstates$}.
  \label{eq:B-eq}
\ee
Note that if $\hat B$ satisfy this equation then the same is true of $\hat B^\dagger$.
It is thus sufficient to consider Hermitian and anti-Hermitian solutions of 
\eqref{eq:B-eq}.

Let us first consider $\hat B$ Hermitian (and non-zero). Then, by letting $s\sdual$
be the projection on the eigenspace of a non-zero eigenvalue, we see that \eqref{eq:B-eq}
does not hold. Next we consider anti-Hermitian $\hat B$. Clearly $B\propto i\idop$
solves \eqref{eq:B-eq}. We claim that these are the only solutions. To see this,
consider a $\hat B$ with two different eigenvalues,
\be
  \hat B |1\rangle = i\lambda_1|1\rangle,\quad \hat B |2\rangle = i\lambda_2|2\rangle,
  \qquad \lambda_1 \neq \lambda_2.
\ee
If we now set
\be
  s = \frac 1 2 (|1\rangle+e^{i\pi/4}|2\rangle)(\langle 1|+e^{-i\pi/4}\langle 2|),\quad
  \sdual = \frac 1 2 (|1\rangle+e^{-i\pi/4}|2\rangle)(\langle 1|+e^{i\pi/4}\langle 2|)
\ee
we find
\be
  \re\tr[s\sdual\hat B] = \frac 1 4 (\lambda_1-\lambda_2) \neq 0
\ee
and the claim follows. This justifies the equivalence $\sim$ in 
\eqref{eq:fancy-gen-observable}.

\section{The von Neumann measurement scheme}
\label{sec:generalized-AAV}
Originally\cite{AAV88}, weak measurements were discussed in the context of a specific 
physical implementation of the measurement process due to von Neumann\cite{Neu32}.
Here we review this formulation of weak measurements and relate it to the results 
of the present paper.

One imagines performing the measurement by coupling the system of interest $\hilb$ to an auxiliary 
meter system $\hilba$. More specifically, let $\hilba = L^2(\mathbb R)$ with the usual operators
$[\hat X,\hat P] = i$. Given an observable $\hat O$ on $\hilb$, we take the interaction between
the system and the meter to be given by the unitary
\be
  \hat U := e^{-i\hat O\hat P}.
\ee
The physical intuition is that the position of the meter ($\hat X$) is shifted by the eigenvalue of 
$\hat O$, but we will see that the situation is more complicated if we postselect on the system.
The initial state of the meter, $\saux^\sigma$, is taken be peaked around $x = 0$, 
with width $\sigma$,
\begin{align}
  \qxpval{\hat X}_{\saux^\sigma} &= 0, &\qxpval{\hat X^2}_{\saux^\sigma} &= \sigma^2.
\end{align}

The expectation value of $\hat X$, after the interaction between the meter and the
system, is simply the expectation value of $\hat O$,
\be
  \tr[(\idop\otimes\hat X)\hat U(s\otimes\saux^\sigma)\hat U^\dagger]
    = \qxpval{\hat O}_s.
  \label{eq:x-expectation}
\ee
When the initial width of meter state is much larger than the eigenvalues of $\hat O$
the measurement becomes weak, with $\sigma^{-1}$ playing the role of the interaction
strength. From the discussion in Section \ref{sec:quant} we then expect
he expectation value of $\hat X$ conditioned on
successful postselection (on the original system) to take the 
form\footnote{When expanding we take $\hat X$ to be of order $\sigma$ and
  $\hat P$ to be of order $\sigma^{-1}$. To make the calculations rigorous, it
  is necessary to add  regularity conditions on $\saux^\sigma$. We omit
  the details.}
\be
  \frac{\tr[(\sdual\otimes\hat X)\hat U(s\otimes\saux^\sigma)\hat U^\dagger]}
    {\tr[(\sdual\otimes\idop)\hat U(s\otimes\saux^\sigma)\hat U^\dagger]}
    = \re \frac{\tr[\sdual \hat A s]}{\tr[\sdual s]}
%      +\qxpval{\{\hat X,\hat P\}}_{\saux^\sigma}\im \frac{\tr[s\sdual \hat O]}{\tr[s\sdual]}
      +O(\sigma^{-1})
\ee
in the weak limit. On one hand it is clear from \eqref{eq:x-expectation} that we must have
$\hat A^R = \hat O$, on the other hand $\hat O$ is the only operator on $\hilb$ in the
game, so we should also have $A^I \propto \hat O$. Indeed, an explicit calculation shows
that\cite{MJP07,DJ12A} 
\be
  \hat A = \hat O - i\qxpval{\{\hat X,\hat P\}}_{\saux^\sigma}\hat O.
  \label{eq:vonN-A}
\ee
The AAV weak value is thus recovered when\cite{Joh04}
\be
  \qxpval{\{\hat X,\hat P\}}_{\saux^\sigma} = 0.
  \label{eq:xp-expectation}
\ee

There are many possible ways to generalize this model of measurement such that $\hat A^I$
does not have to be proportional to $\hat O$. One possibility is to replace 
$\hat U \to \hat U^\sigma$,
\be
  \hat U^\sigma := e^{i\frac{\sigma^{-2}}{2}\hat B\hat X}e^{-i\hat O\hat P}.
\ee
Here $\hat B$ is an arbitrary Hermitian operator on $\hilb$. In this generalized
model \eqref{eq:x-expectation} still holds (for any finite $\sigma$), but
now the conditional expectation value is
\be
  \frac{\tr[(\sdual\otimes\hat X)\hat U^\sigma(s\otimes\saux^\sigma)(\hat U^\sigma)^\dagger]}
    {\tr[(\sdual\otimes\idop)\hat U^\sigma(s\otimes\saux^\sigma)(\hat U^\sigma)^\dagger]}
    = \re \frac{\tr[\sdual \hat A' s]}{\tr[\sdual s]}
      +O(\sigma^{-1}),
\ee
with
\be
  \hat A' = \hat O + i(\hat B-\qxpval{\{\hat X,\hat P\}}_{\saux^\sigma}\hat O).
\ee

Before we turn to disturbance, let us briefly examine how the meter system
is affected by the interaction. The probability distribution of the meter
position $\hat X$ is initially
\be
  P_i(x) := \qxpval{\hat\Pi_x}_{\saux^\sigma},\qquad \Pi_x := |x\rangle\langle x|.
\ee
After the interaction and postselection of the system it becomes
\begin{multline}
  P_f(x) := \frac{\tr[(\sdual\otimes\hat \Pi_x)\hat U^\sigma
                    (s\otimes\saux^\sigma)(\hat U^\sigma)^\dagger]}
    {\tr[(\sdual\otimes\idop)\hat U^\sigma(s\otimes\saux^\sigma)(\hat U^\sigma)^\dagger]}
   = P_i(x)
    - \left(\re \frac{\tr[\sdual \hat O s]}{\tr[\sdual s]}\right)
      \partial_x P_i(x) \\
    + \left(\im \frac{\tr[\sdual \hat O s]}{\tr[\sdual s]}\right)
       \qxpval{\{\hat\Pi_x,\hat P\}}_{\saux^\sigma}
    - \left(\im \frac{\tr[\sdual \hat B s]}{\tr[\sdual s]}\right)
      \sigma^{-2}x P_i(x)
    + O(\sigma^{-3})
\end{multline}
to lowest non-trivial order. With no postselection only the two first terms contribute,
and we see that the meter (distribution) is simply translated, in accordance with
the  physical intuition. However, once we postselect this picture is in general ruined by
the additional terms, even if $\hat B$ is zero (i.e.~in the original von Neumann model).
This shows that one should be careful about applying intuition to the quantum measurement
process, even for simple models like von Neumann's.

The state of the system after the weak measurement is
\be
  M_?^\sigma(s) = \tr_{\hilba}[\hat U^\sigma(s\otimes\saux^\sigma)(\hat U^\sigma)^\dagger]
  = s-i\qxpval{\hat P}_{\saux^\sigma}[\hat O,s]+O(\sigma^{-2}).
\ee
We thus conclude that the measurement is non-disturbing in the weak sense iff
$\qxpval{\hat P} = 0$. Note that this condition does not put any constraints on
$\qxpval{\{\hat X,\hat P\}}$ or $\hat B$. The average survival probability is
\be
   \int\ud\psi\, \tr \left[M_?^\sigma(|\psi\rangle\langle\psi|) |\psi\rangle\langle\psi|\right]
    = 1-\frac{\sigma^{-2}}{d(d+1)}\distmeas
    + O(\sigma^{-3}), 
\ee
with
\be
  \distmeas = \sigma^2\qxpval{\hat P^2}_{\saux^\sigma}\distmeasf(\hat O)
    + \frac 1 4 \distmeasf(\hat B)
    - \frac 1 2 \qxpval{\{\hat X,\hat P\}}_{\saux^\sigma}(d\tr[\hat O\hat B]
        -\tr[\hat O]\tr[\hat B]).
\ee
This expression is, for fixed $\hat O$, bounded from below. In fact,
\begin{align}
  \distmeas &= \left(\sigma^2\qxpval{\hat P^2}_{\saux^\sigma}
               -\frac 1 4 \qxpval{\{\hat X,\hat P\}}_{\saux^\sigma}\right)\distmeasf(\hat O)
              + \frac 1 4 \distmeasf\left(\qxpval{\{\hat X,\hat P\}}_{\saux^\sigma} \hat O
                - \hat B\right)\nonumber\\
     &\geq \left(\sigma^2\sigma_p^2-\frac 1 4 \qxpval{\{\hat X,\hat P\}}_{\saux^\sigma}\right)
               \distmeasf(\hat O)\nonumber\\
     &\geq \frac 1 4 \distmeasf(\hat O),
     \label{eq:disturbance-bound}
\end{align}
where the last inequality is the Schr\"{o}dinger uncertainty relation.
Note that for non-trivial observables $\hat O$, the inequality implies that
$\distmeas$ is strictly larger that zero.
The situation considered by AAV\cite{AAV88} corresponds to
$\qxpval{\{\hat X,\hat P\}} = 0$,
$\hat B = 0$ and $\qxpval{\hat P^2} = \sigma^{-2}/4$ which implies
that $\distmeas = \distmeasf(\hat O)/4$.
We see that the AAV measurement procedure minimizes the value of $\distmeas$, in
accordance with the general result of Section \ref{sec:uniqueness}.
%\section{Disturbance and Lindblad superoperators}
%
%\be
%  \khat_{m,n}^\lambda s (\khat_{m,n}^\lambda)^\dagger
%    = \frac 1 2 ((\khat_{m,n}^\lambda)^\dagger \khat_{m,n}^\lambda s + s (\khat_{m,n}^\lambda)^\dagger \khat_{m,n}^\lambda)
%    + \lindblad(\khat_{m,n})s
%\ee
%
%\be
%  \lindblad(\khat_{m,n})s := \khat_{m,n}^\lambda s (\khat_{m,n}^\lambda)^\dagger
%    - \frac 1 2 ((\khat_{m,n}^\lambda)^\dagger \khat_{m,n}^\lambda s + s (\khat_{m,n}^\lambda)^\dagger \khat_{m,n}^\lambda)
%\ee
%
%\be
%  M_?^\lambda(s) = s + \sum_{m,n}\lindblad(\khat_{m,n})s
%\ee

\section{Minimally disturbing measurements}
\label{sec:uniqueness-proof}
We want to characterize the minimally disturbing (in the sense of having the smallest
$\distmeas$ as defined in Sec.~\ref{sec:uniqueness}) generalized weak measurements.
More concretely, consider the collection $\measmnts_{\hat O}$ of 
weak measurements\footnote{As in Sec.~\ref{sec:quant} we consider a weak measurement procedure to
  be specified by constants $A_m$ and $K^0_{m,n}$, and operators $\delta\khat_{m,n}$.}
measuring a fixed observable without postselection, i.e.~such that $\hat A^R = \hat O$.
As a first guess one might try to minimize $\distmeas$ on $\measmnts_{\hat O}$, 
but this fails because the simple rescaling 
\be
  A_m \to \epsilon^{-1}A_m,\qquad K^0_{m,n}\to K^0_{m,n},\qquad
  \delta\khat_{m,n}\to \epsilon \delta\khat_{m,n}
  \label{eq:rescaling}
\ee 
shows that there are elements of $\measmnts_{\hat O}$ with arbitrarily small $\distmeas$.

Let $N$ be the number of measurement outcomes, which we consider fixed.
A natural choice is to  consider the subset $\measmnts_{\hat O,A^*} \subset \measmnts_{\hat O}$ where 
 $(A_{m=1},\ldots,A_{m=N})$ is constrained to belong to some \emph{compact} 
set $A^* \subset {\mathbb R}^N$. 
In this way we avoid the problem of the rescaling \eqref{eq:rescaling}, since compact sets
are bounded. The exact nature of the set $A^*$ is not important, except that $\measmnts_{\hat O,A^*}$
should be non-empty.
Unfortunately  $\measmnts_{\hat O,A^*}$ 
is not compact, so the existence of a minimal elements is still not obvious.
To remedy this problem we define a better behaved subset 
$\measmnts_{\hat O,A^*}'\subset\measmnts_{\hat O,A^*}$ such that to each element $x$ of
$\measmnts_{\hat O,A^*}$ there corresponds an element $y$ of $\measmnts_{\hat O,A^*}'$
with $\distmeas(y) \leq \distmeas(x)$. It is then clear that a minimal element of
$\measmnts_{\hat O,A^*}'$ is also a minimal element of $\measmnts_{\hat O,A^*}$.
What we will show is:

\vspace{3mm}
  \emph{If the set $\measmnts_{\hat O,A^*}$ is non-empty then it contains elements minimal
  with respect to $\distmeas$. Furthermore, these minimal elements satisfy $\hat A\sim \hat O$
  which implies that the weak values are given by the AAV formula.}
\vspace{3mm}

Let us first note that setting all $\delta\khat_{m,n}^I = 0$  decreases 
$\distmeas$ (see Eq.~\eqref{eq:distmeas-split}) and does not change $\hat A^R$.
We can thus restrict $\measmnts_{\hat O,A^*}'$ to having Hermitian $\delta\khat_{m,n}$.
We can also restrict to having only one Kraus operator per measurement outcome.
To see this fix an $m$ and consider the contribution $\distmeas_m$ to $\distmeas$ from this 
outcome.
We then have the inequalities
\begin{align}
  P^0_m \distmeas_m &= 
    \left(\sum_n (K^0_{m,n})^2\right)\left(\sum_n \tr[\delta\khat_{m,n}^2]\right)\\
  &\geq \left| \sum_n K^0_{m,n}\sqrt{\tr [\delta\khat_{m,n}^2]}\right|^2\\
  &\geq \tr\left[\left(\sum_n K^0_{m,n}\delta\khat_{m,n}\right)^2\right]\\
  &= \tr[\delta\bar K_m^2],
\end{align}
that is
\be
  \distmeas_m \geq \frac{\tr[\delta\bar K_m^2]}{P^0_m}.
\ee
But this shows that replacing $\delta\khat_{m,n}$ by a single operator 
$\delta\khat_m$ given by
\be
  \delta\khat_{m} = \frac{\sum_{n}K^0_{m,n}\delta\khat_{m,n}}{\sqrt{P^0_m}}
\ee 
(along with $K^0_{m,n} \to K^0_m = \sqrt{P^0_m}$) decreases $\distmeas_m$
and thus $\distmeas$.

To simplify matters slightly let us assume $\tr \hat O = 0$ for now. We note
that $f(\delta\khat+c\idop) = f(\delta\khat)$ for any $c\in\mathbb R$. It 
follows that the replacement
\be
  \delta\khat_m \to \delta\khat_m-(d^{-1}\tr[\delta\khat_m])\idop
  \label{eq:khat-shift}
\ee
leaves $\distmeas$ invariant. This allows us to restrict $\measmnts_{\hat O,A^*}'$
to measurements with $\tr[\delta\khat_m] = 0$.

On elements of $\measmnts_{\hat O,A^*}'$ $\distmeas$ is given by
\be
  \distmeas = d\sum_m \tr[\delta\khat_m^2].
\ee
For sufficiently big $C$ the set
\be
  \measmnts_{\hat O,A^*}'' := \{x\in \measmnts_{\hat O,A^*}' \mid \distmeas(x) \leq C\}
\ee
is seen to be non-empty and compact (here the compactness of $A^*$ is needed),
and a minimal element of $\measmnts_{\hat O,A^*}''$
is also minimal in $\measmnts_{\hat O,A^*}'$ and hence in $\measmnts_{\hat O,A^*}$. We have
thus shown that there are minimally disturbing measurements. The general case 
of $\tr \hat O \neq 0$ is easily reduced to the case we have covered by shifting by
the identity (similar to Eq.~\eqref{eq:khat-shift}).

Using that $f(\delta\khat_{m,n}^I) = 0$ iff $\delta\khat_{m,n}$ is proportional
to the identity and Eq.~\eqref{eq:distmeas-split} it is clear that for minimally
disturbing
measurements in $\measmnts_{\hat O,A^*}$ we must have $\hat A \sim \hat O$, which
is what we wanted to show. We leave a more thorough characterization of the 
minimally disturbing measurements to further work.

\end{document}